\documentclass[]{interact}

\usepackage{epstopdf}
\usepackage[caption=false]{subfig}

\usepackage[numbers,sort&compress]{natbib}
\bibpunct[, ]{[}{]}{,}{n}{,}{,}
\renewcommand\bibfont{\fontsize{10}{12}\selectfont}
\makeatletter
\def\NAT@def@citea{\def\@citea{\NAT@separator}}
\makeatother

\theoremstyle{plain}

\theoremstyle{definition}

\theoremstyle{remark}

\begin{document}

\articletype{}

\title{Numerical study of the transverse localization of waves in one-dimensional lattices with randomly distributed gain and loss: Effect of disorder correlations}

\author{
\name{Ba Phi Nguyen\textsuperscript{a}\thanks{CONTACT B.~P. Nguyen. Email: nguyenbaphi@muce.edu.vn}, Thi Kim Thoa Lieu\textsuperscript{b} and Kihong Kim\textsuperscript{c}\thanks{CONTACT K. Kim. Email: khkim@ajou.ac.kr}}
\affil{\textsuperscript{a}Department of Basic Sciences, Mientrung University of Civil Engineering, Tuy Hoa 620000, Vietnam; \textsuperscript{b}Department of Physics, Quy Nhon University, Quy Nhon 820000, Vietnam;
\textsuperscript{c}Department of Energy Systems Research and Department of Physics, Ajou University, Suwon 16499, Korea
}}

\maketitle

\begin{abstract}
We study numerically the effects of short- and long-range correlations on the localization properties of the eigenstates
in a one-dimensional disordered lattice characterized by a random non-Hermitian Hamiltonian,
where the imaginary part of the on-site potential is random.
We calculate participation number versus strengths of disorder and correlation.
In the short-ranged case and when the correlation length is sufficiently small,
we find that there exists a critical value of the disorder strength,
below which enhancement and above which suppression of localization occurs as the correlation length increases.
In the region where the correlation length is larger, localization is suppressed in all cases.
A similar behavior is obtained for long-range correlations as the disorder strength and the correlation exponent are varied.
Unlike in the case of a long-range correlated real random potential, no signature of the localization transition is found in a long-range correlated imaginary random potential. In the region where localization is enhanced
in the presence of long-range correlations,
we find that the enhancement occurs in the whole energy band, but is strongest near the band center.
In addition, we find that the anomalous localization enhancement effect occurs near the band center in the long-range correlated case.
\end{abstract}


\section{Introduction}

Though Anderson localization of quantum particles and classical waves has been studied extensively for over six decades, it is still
attracting the interest of many researchers in the different areas of physics \cite{And,Sheng,Abrahams,mod,gre,seg,seg2,kk0,kk1}.
New aspects of localization phenomena have been discovered
continually in diverse systems characterized by
different types of wave equations or random potentials. Among them, the combined influence of the spatial correlation and the non-Hermiticity
of the random potential is the main focus of this study.

Recently, much attention has been given to the unique physical phenomena arising in the systems described
by non-Hermitian Hamiltonians \cite{Moise,gana1,gana2,Hat,Bender,Rud,Ruter,Eic,Zue}.
Examples of this kind can be found in a wide range of problems in all areas of physics.
Non-Hermitian quantum mechanics has been developed to describe open quantum systems and the resonance phenomena occurring in them
\cite{Moise,Bender,Rotter,Soko,Eleuch3}.
Non-Hermitian Hamiltonians have also been used to study the dynamical behavior of various classical mechanical systems \cite{Hat,shnerb}.
An extensively studied case is the propagation of light in complex optical systems with engineered loss and/or gain,
where the non-Hermiticity enters through the imaginary part of the refractive index \cite{gana1,gana2,Ruter,Eic,Zue}.
This area has attracted much interest from researchers and even a new field called non-Hermitian optics has emerged \cite{gana2}.
In non-Hermitian systems with balanced gain and loss, peculiar degeneracies of the spectrum called exceptional points can appear \cite{miri}.
Since the physical properties can be varied extremely sensitively near the exceptional point,
many theoretical and experimental studies have been carried out to explore the characteristics of exceptional-point singularities
in non-Hermitian optical, electronic, acoustical,
and atomic systems and use the results to develop various sensors such as hypersensitive optical gyroscopes and wireless readouts \cite{sak,xiao,lai,hok,dong,dom}.
In the special case where the complex potential satisfies the combined parity and time-reversal symmetry, the system has
a real-valued spectrum and can display interesting topological properties \cite{gana1,gana2,Bender,Ruter,Eic,sak,xiao,Ozawa}.

The effective Hamiltonian of an open quantum system is non-Hermitian \cite{Moise,Eleuch3}.
In this case, the non-Hermiticity arises from the interaction of the quantum system with the environment,
which can consist of either the measuring device that connects to the system or the natural environment into which the system is embedded.
Open quantum systems are found in many areas of physics including optics, atomic and nuclear physics, mesoscopic physics, and biophysics
and the theoretical study of their properties often employs various methods of non-Hermitian quantum mechanics \cite{Moise,Soko}.

During the recent few decades, the influence of spatial disorder correlations on the localization properties has been studied
extensively \cite{Izrailev3,Dunlap,Bellani,Moura,Izrailev1,Kantelhardt,Kuhl1,Carpena,Shima,Kaya1,Nishino,Garcia,Gong,Croy,Nguyen2,Albrecht,Petersen,choi,Hilke,lima}. It has been demonstrated both theoretically and experimentally that
some special types of short-range correlated disorder can produce extended eigenstates even in
one dimension \cite{Dunlap,Bellani,choi}.
More interestingly, it was predicted that the introduction of long-range correlations into a disordered system would suppress localization and give rise to a continuum of extended states in some cases \cite{Moura,Izrailev1}.
Although this prediction caused some controversy \cite{Moura,Kantelhardt},
it was experimentally verified later by a measurement of microwave transmission spectra
through a single-mode waveguide with artificially introduced correlated disorder \cite{Kuhl1}.
In spite of a large number of publications devoted to this topic,
the problem of understanding the effects of short- and long-range correlations on
Anderson localization has not been completely resolved yet. An important and elementary question is
whether the presence of spatial correlations will always suppress localization as stated in some of the previous studies.

Many of the theoretical studies quoted above are based on the numerical solution of discrete lattice models. In this regard,
it is highly desirable to obtain
accurate analytical results on the localization in a correlated random potential \cite{kk0,kk1,Hilke,Eleuch2,Eleuch1}. For instance, in \cite{Eleuch2}, it
has been shown that the localization properties of a disordered wave equation can be obtained at all values of the disorder strength
and the correlation length by using the disorder average of an approximate nonlinear wave equation.

Another interesting direction in the study of localization is to consider the effects of non-Hermitian
random potentials, which include imaginary scalar and vector potentials
and ${\cal PT}$-symmetric complex potentials \cite{Hat,Asa,Silvestrov,Kal,Jov,Eic,Basiri,Vaz,Mej,Kar,ba}.
It has been reported that these systems show many interesting localization properties
such as a transition from a real to a complex spectrum \cite{Hat}.
Recently, a one-dimensional (1D) non-Hermitian lattice model with randomness
only in the imaginary part of the on-site potential has been studied,
with the result that the eigenstates of such a model are localized, but the properties of localization
are qualitatively different from those of the usual Anderson model \cite{Basiri,Nguyen1}.
To the best of our knowledge, however, the combined effect of correlations and non-Hermiticity on localization has not been
studied in detail.

It is difficult to observe Anderson localization in electronic systems due to the existence of
Coulomb and electron-phonon interactions. However, the localization of electromagnetic waves in random dielectric media
has been observed experimentally, though the difficulty to observe it
in three dimensions remains to be resolved \cite{seg,seg2,Wiersma1,Chabanov,Schwartz,sperl}.
One important application of Anderson localization in optical systems is to develop random lasers \cite{law,Cao,Noginov,Wiersma2,abaie},
where an optical cavity is formed not by mirrors but by multiple scattering in a random gain medium.
Such a medium, where gain and loss are distributed randomly, can be considered as a typical example of a non-Hermitian random medium.
To achieve high efficiency output from a random laser, it is crucial to produce strongly localized states.
There is a common point of view that the strongest localization is obtained for strongly disordered uncorrelated potentials,
which is based on the prediction that correlations suppress, rather than enhance, localization.
In some specific situations \cite{Kuhl2,Zhao,Deng,Nezhad}, however, a strong decrease of the localization length and therefore
a strong enhancement of localization has been observed, when
long-range correlated disorder is introduced into the system.
These findings can provide an insight into achieving correlation-induced localization enhancement
in other disordered wave systems.
Apart from random lasers, the transverse Anderson localization of light has already
found device-level applications in biological and medical imaging \cite{Ka1,Ka2,Ka3}.

In the present paper, we study the effects of short- and long-range disorder correlations
on the localization properties of the eigenstates in a 1D disordered lattice characterized by a random non-Hermitian Hamiltonian, where
the imaginary part of the on-site potential is a correlated random function of the position.
We calculate the participation number, which estimates the degree of spatial extension or localization of eigenstates, as a function of other parameters. We find that the presence of both short- and long-range correlations in the imaginary random potential
can cause strong enhancement of the localization effects in some parameter regimes.
This effect can be important in applications such as random lasers, where strongly localized states are desired to
achieve high efficiency.

The rest of this paper is organized as follows. In Section~\ref{sec:model}, we describe the 1D disordered lattice model characterized by a random non-Hermitian Hamiltonian within the nearest-neighbor tight-binding approximation. The numerical calculation method and the physical quantity of main interest are also described. In Section~\ref{sec:result}, we present our numerical results and discussions. Finally, in Section~\ref{sec:con}, we conclude the paper.

\section{Method}
\label{sec:model}

\subsection{Model}

We consider a one-dimensional array of $N$ weakly coupled optical waveguides. We assume that light is transported between adjacent waveguides
through optical tunneling with a uniform tunneling amplitude $V$.
If we consider only the nearest-neighbor coupling, then the wave propagation in such a system can be described by
a set of discrete linear Schr\"{o}dinger equations for the field amplitudes $C_{n}$:
\begin{eqnarray}
i \frac{dC_{n}}{dz}=-V\left(C_{n-1}+C_{n+1}\right) + \epsilon_{n}C_{n},
\label{equation1}
\end{eqnarray}
where $z$ is the paraxial coordinate, $\epsilon_{n}$ is the on-site potential and $n$ ($=1,2,\cdots,N$) is the waveguide index.
The stationary solutions of this equation can be written as $C_{n}(z)=\psi_{n}e^{-iEz}$, where $E$ is the energy of an eigenstate. Then we obtain
the stationary discrete Schr\"{o}dinger equation for the amplitude $\psi_{n}$:
\begin{eqnarray}
 E{\psi_{n}}=-V\left(\psi_{n-1}+\psi_{n+1}\right)+\epsilon_{n}\psi_{n}.
\label{equation2}
\end{eqnarray}

In order to obtain the eigenstates and the corresponding eigenvalues, we need to solve Equation~(\ref{equation2}) numerically using suitable
boundary conditions. In this work, we use the fixed boundary conditions $\psi_{0}=\psi_{N+1}=0$.
We also consider the on-site potential $\epsilon_{n}$ to be a complex number given by
\begin{eqnarray}
\epsilon_{n}=\epsilon_{n}^{\rm R}+i\epsilon_{n}^{\rm I},
\label{equation3}
\end{eqnarray}
where the real part $\epsilon_{n}^{\rm R}$ is assumed to be the same for all lattice sites (below we put $\epsilon_{n}^{\rm R}=0$ for simplicity), whereas the imaginary part $\epsilon_{n}^{\rm I}$ is a random variable which is either short-range-correlated or long-range-correlated.

\subsection{Participation number}

In this paper, we characterize the localization properties in terms of the participation number.
For the $k$-th eigenstate $(\psi_{1}^{(k)},\psi_{2}^{(k)},\cdots,\psi_{N}^{(k)})^{\rm T}$ with the eigenvalue $E_{k}$, the participation number $P(E_{k})$ is defined by \cite{Thouless}
\begin{eqnarray}
P(E_{k})=\frac{\left(\sum_{n=1}^{N}\big\vert\psi_{n}^{(k)}\big\vert^2\right)^2}{\sum_{n=1}^{N}\big\vert\psi_{n}^{(k)}\big\vert^4},
\label{equation4}
\end{eqnarray}
which estimates the number of lattice sites to which the $k$-th eigenstate extends.

The participation number measures the degree of spatial extension or localization of eigenstates. In the extended regime,
$P$ increases monotonically with increasing the system size $N$. The most extended state which spreads over the entire system uniformly corresponds to $P=N$. In the localized regime, $P$ is much smaller and converges to a constant value as $N\to \infty$. The most strongly localized state corresponds to $P=1$. Therefore, the participation number is bounded in the range $1\le P \le N$.

The discrete Schr\"odinger equation, Equation~(\ref{equation1}), which describes the light propagation in optical systems, is similar to the equation describing the transport of noninteracting electrons in electronic systems. The difference is that the evolution coordinate in the optical case is the paraxial propagation distance, whereas it is replaced by the time in the electronic case.

\section{Numerical results}
\label{sec:result}

In our numerical calculations, all energy quantities are measured in the unit of $V$, which we set equal to 1 with no loss of generality. The participation number $P$ was obtained by averaging over the eigenstates with eigenvalues in a small interval around a fixed ${\rm Re}(E)$ and ensemble averaging over 10000 different disorder configurations. In a recent work \cite{Nguyen1}, we studied the detailed localization properties
of the systems where only the imaginary part of the on-site potential is an uncorrelated random variable of the position. In the present study, we will generalize this work further and investigate the role of short-range and long-range disorder correlations on the localization in the systems where the imaginary part of the on-site potential is random. Although our main focus is to consider the effect of correlations on the localization
arising from random variations in the imaginary part of the on-site potential, we will also reproduce
some results for the case of the real random potential.

\subsection{Short-range correlations}

\begin{figure}
\centering
\includegraphics[width=10cm]{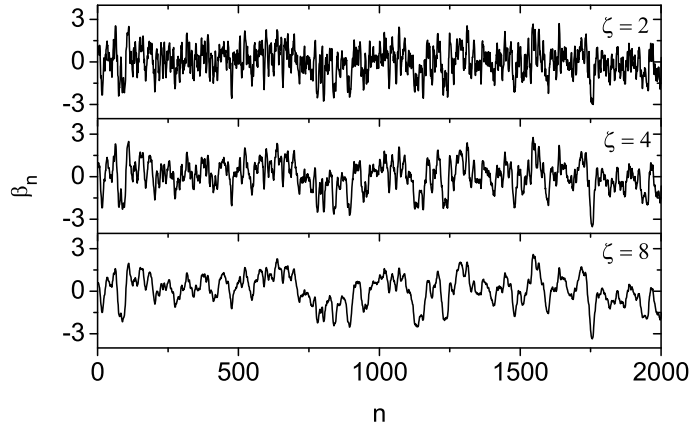}
\caption{Typical short-range correlated random configurations of $\beta_n$, when $\sigma=0.1$ and $\zeta=2$, 4, 8.}
\label{fig:1}
\end{figure}

Short-range correlated random sequences can be generated using several different methods \cite{Izrailev2,Kaya2,Izrailev3,Sales,Herrera}. In this study, we follow the procedure described in \cite{Sales} and \cite{Herrera}. We first generate a large number of independent random numbers
$\{\eta_m\}$ uniformly distributed in the interval $[-1/2,1/2]$. Then we
calculate the sequence $\{\tilde\beta_{n}\}$ ($n=1,2,\cdots,N$) using
\begin{eqnarray}
\tilde\beta_{n}=\sum_{m}\eta_{m}e^{-\vert n-m\vert/\zeta},
\label{equation5}
\end{eqnarray}
where the parameter $\zeta$ is the disorder correlation length. When $\vert n-m\vert$ is smaller than $\zeta$,
the two random variables $\tilde\beta_n$ and $\tilde\beta_m$ are no longer independent.

In order to obtain a suitable random on-site potential, we need to normalize the sequence $\{\tilde\beta_{n}\}$ using
\begin{eqnarray}
{\beta}_n=\sigma\frac{\tilde\beta_{n}}{\sqrt{\langle{\tilde{\beta_{n}}^{2}} \rangle}},
\label{equation6}
\end{eqnarray}
where $\langle\cdots\rangle$ denotes averaging over a large number of distinct disorder configurations. The parameter $\sigma$ measures the strength of disorder. In \cite{Sales}, it was shown numerically that the two-point correlation function $\langle {\beta}_i{\beta}_j\rangle$ exhibits an exponential decay. This exponential decay is not completely smooth due to the presence of the prefactor preceding the exponential term in the two-point correlation function \cite{Herrera}. However, this does not affect the short-range correlated nature of the disorder distribution. In the limit where $\zeta\to 0$, we recover the case of an uncorrelated random potential. For a finite $\zeta$, a disorder distribution with short-range correlations is generated. In Figure~\ref{fig:1}, we show an example of typical configurations of $\beta_n$ corresponding to $\sigma=0.1$ and
$\zeta=2$, 4, 8.

\begin{figure}
\centering
\includegraphics[width=10cm]{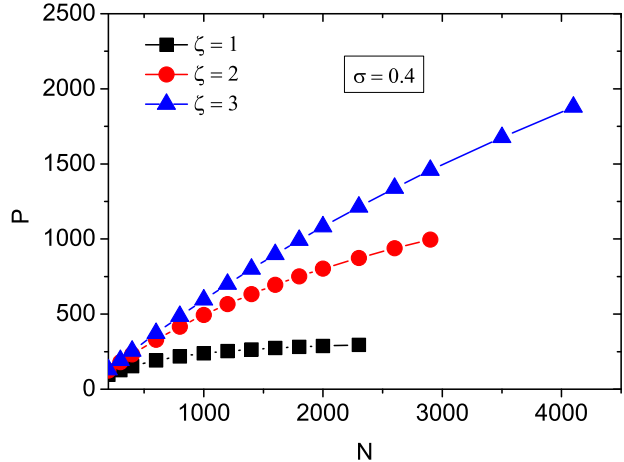}
\caption{Participation number $P$ plotted versus system size $N$,
when the short-range correlated random on-site potential is real-valued with zero mean.
The strength of disorder is fixed to $\sigma=0.4$ and the disorder correlation length is chosen to be $\zeta=1$, 2 and 3.
In this parameter region, Anderson localization is suppressed due to short-range correlation.}
\label{fig:a}
\end{figure}

We first consider the case of a real random potential, where $\epsilon_{n}={\beta}_n$. In Figure~\ref{fig:a}, we plot the participation number $P$ as a function of the system size $N$ for several values of the disorder correlation length $\zeta$ ($=1,2,3$), when the strength of disorder $\sigma$ is fixed
to 0.4. A small energy interval around the band center such that $E\in [-0.1,0.1]$ has been considered. We find that for a fixed system size, $P$ increases as $\zeta$ increases, which implies that Anderson localization is suppressed due to short-range correlation in this parameter region.
This behavior is in good agreement with the results of previous studies \cite{Dunlap,Bellani,Izrailev3}.

\begin{figure}
\centering
\includegraphics[width=10cm]{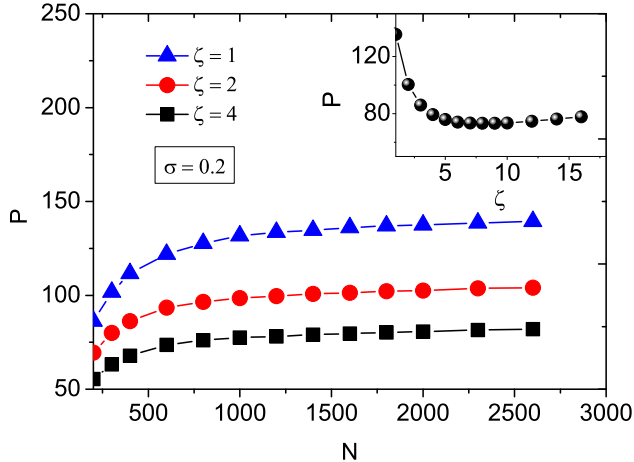}
\caption{Participation number $P$ versus system size $N$,
when the short-range correlated random on-site potential is imaginary-valued with zero mean.
The strength of disorder is fixed to $\sigma=0.2$ and the disorder correlation length is chosen to be $\zeta=1$, 2 and 4.
A small energy interval around the band center such that $\rm Re(E)\in [-0.1,0.1]$ has been considered.
In this parameter region, Anderson localization is enhanced due to the short-range correlation of the imaginary random potential.
In the inset, $P$ is plotted versus $\zeta$ when the system size is fixed to $N=2000$.}
\label{fig:b}
\end{figure}

Next, we consider the case of a pure imaginary random potential, where $\epsilon_{n}=i{\beta}_n$.
Since ${\beta}_n$ can take both positive and negative values with equal probability,
we are dealing with a model with randomly distributed gain and loss.
In the case of an uncorrelated disorder corresponding to $\zeta=0$, it has been demonstrated previously that
exponentially localized states occur in the presence of random fluctuations in the imaginary part of the on-site potential \cite{Basiri, Nguyen1}.
In Figure~\ref{fig:b}, we plot $P$ versus $N$ for several values of $\zeta$ ($=1,2,4$), when the strength of disorder is fixed
to $\sigma=0.2$. A small energy interval around the band center such that ${\rm Re}(E)\in [-0.1,0.1]$ has been considered.
We find that the large $N$ limiting values of $P$ decreases as $\zeta$ increases, which implies that Anderson localization is enhanced due to short-range correlation in this parameter region. This behavior is opposite to that shown in Figure~\ref{fig:a}, though the parameter values used in the two
calculations are similar.
In the inset of Figure~\ref{fig:b}, we plot $P$ versus $\zeta$ for fixed values of $N$ ($=2000$) and and $\sigma$ ($=0.2$). We find that as $\zeta$ increases from zero,
$P$ decreases initially, reaches a minimum and then increases slowly.

\begin{figure}
\centering
\includegraphics[width=10cm]{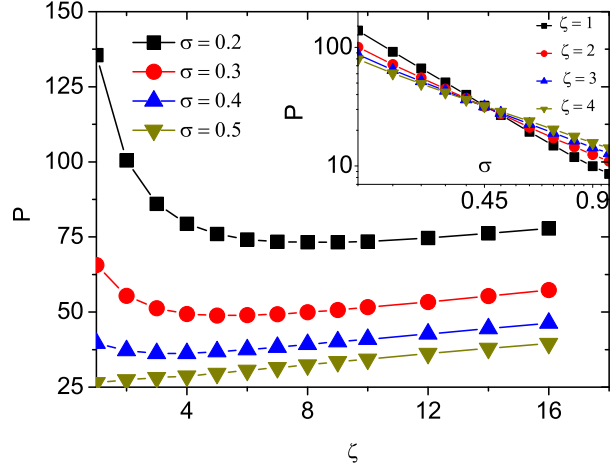}
\caption{Participation number $P$ versus correlation length $\zeta$,
when the short-range correlated random on-site potential is imaginary-valued with zero mean.
The system size is fixed to $N=2000$ and the strength of disorder is chosen to be $\sigma=0.2$, 0.3, 0.4 and 0.5.
In the weak correlation regime where $\zeta\leq 4$, there exists a critical value $\sigma_{c}\simeq 0.45$ above (below) which the localization is suppressed (enhanced) with increasing $\zeta$, as demonstrated in the inset.}
\label{fig:c}
\end{figure}

The non-monotonic dependence of $P$ on the correlation length is considered in more details in Figure~\ref{fig:c},
where $P$ is plotted versus $\zeta$ for several different values of the disorder strength $\sigma$.
The system size is fixed to $N=2000$, which ensures that the obtained results do not come from the finite-size effects.
We find that when $\sigma=0.2$, 0.3 and 0.4, $P$ shows a non-monotonic dependence on $\zeta$, while when $\sigma=0.5$, it increases monotonically
as $\zeta$ increases from 1. This implies that in the region where the correlation length is sufficiently large,
the localization is suppressed due to short-range disorder correlations of the imaginary random potential.
On the other hand, in the weak correlation regime where $\zeta\leq 4$, there exists a critical value $\sigma_{c}\simeq 0.45$ above (below) which the localization is suppressed (enhanced) with increasing $\zeta$, as demonstrated clearly also in the inset of Figure~\ref{fig:c}.

\subsection{Long-range correlations}

\begin{figure}
\centering
\includegraphics[width=10cm]{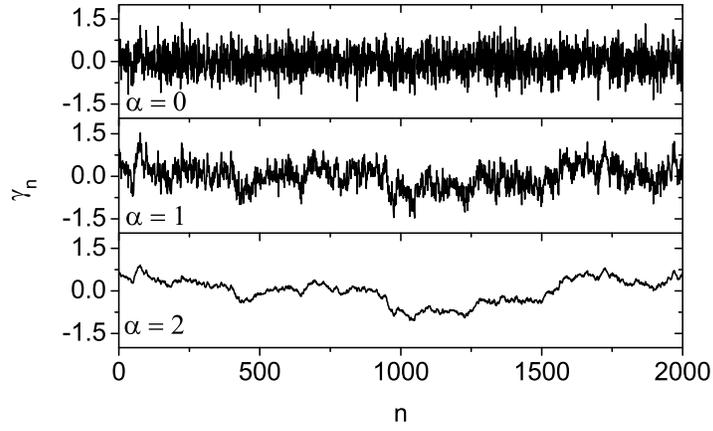}
\caption{Typical random configurations of $\gamma_n$, when $W$ is 1.5 and the long-range correlation exponent $\alpha$
is 0, 1, 2.}
\label{fig:2}
\end{figure}

There exist two main methods for generating long-range correlated random sequences, which are the fractional Brownian motion trace \cite{Greis,Rangarajan} and the Fourier filtering method \cite{Prakash,Makse}. In this study, we apply the former method, which has also been employed by de Moura and Lyra \cite{Moura} and by two of us \cite{Nguyen2}, to generate the sequence $\{\tilde\gamma_{n}\}$ defined by
\begin{eqnarray}
\tilde\gamma_{n}=\sum_{k=1}^{N/2}\left[k^{-\alpha}\left(\frac{2\pi}{N} \right)^{(1-\alpha)}\right]^{1/2}\cos \left(\frac{2\pi n k}{N} +\phi_{k} \right),
\label{equation7}
\end{eqnarray}
where the number of sites $N$ is an even number and $\phi_{k}$'s are $N/2$ random phases uniformly distributed in the interval $[0,2\pi]$.  The Fourier transform of the two-point correlation function $\langle \tilde\gamma_{i}\tilde\gamma_{j}\rangle$, $S(k)$, is proportional to a power-law spectrum $k^{-\alpha}$. The exponent $\alpha$ determines the roughness of potential landscapes and characterizes the long-range correlation strength.
When $\alpha$ is zero, we recover an uncorrelated random sequence. In order to study the dependence on the strength of disorder properly, we need to normalize the generated sequence $\{\tilde\gamma_{n}\}$  by multiplying a suitable normalization constant \cite{Kaya1}
\begin{eqnarray}
{\gamma}_n=\frac{W}{\sqrt{12}}\frac{\tilde\gamma_{n}}{C},
\label{equation8}
\end{eqnarray}
where $W$ measures the strength of disorder and the constant $C$ is chosen such that
\begin{eqnarray}
\sqrt{ \langle {\gamma_n}^2 \rangle - {\langle \gamma_n\rangle}^2}=\frac{W}{\sqrt{12}}.
\label{equation9}
\end{eqnarray}
In Figure~\ref{fig:2}, we show an example of typical configurations of $\gamma_n$ corresponding to $W=1.5$ and
$\alpha=0$, 1, 2.

\begin{figure}
\centering
\includegraphics[width=10cm]{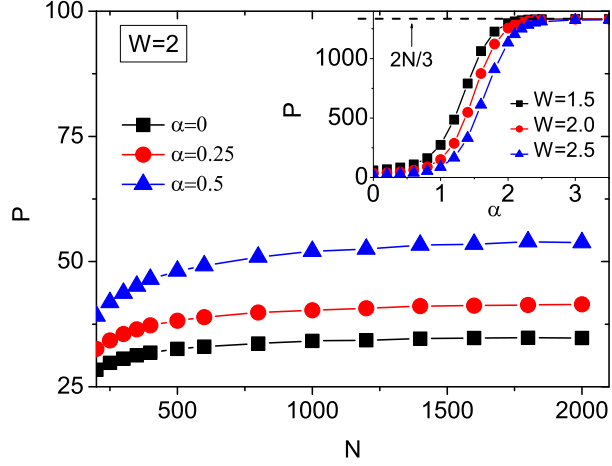}
\caption{Participation number $P$ versus system size $N$,
when the long-range correlated random on-site potential is real-valued with zero mean.
The strength of disorder is fixed to $W=2$ and the correlation strength is chosen to be $\alpha=0$, 0.25 and 0.5.
In this parameter region, Anderson localization is suppressed due to the long-range correlation of the real random potential.
In the inset, $P$ is plotted versus $\alpha$ when $N=2000$ and $W=1.5$, 2 and 2.5. It is seen that there is a correlation-induced localization-delocalization transition at a critical value $\alpha_{c}$, the value of which depends on the disorder strength $W$.}
\label{fig:d}
\end{figure}

Similarly to the case of short-range correlation, we first consider the case of a long-range correlated real random potential. Although this case was already considered extensively \cite{Moura,Izrailev1,Kuhl1,Carpena,Shima,Kaya1,Nishino,Garcia}, we present it here in a slightly different manner. In Figure~\ref{fig:d}, we show the participation number $P$ as a function of the system size $N$ for different values of the correlation strength $\alpha$ ($=0,0.25,0.5$),
when the disorder strength $W$ is fixed to 2. Our numerical results show clearly that the presence of long-range correlations gives rise to a strong suppression of localization. When the correlation exponent $\alpha$ is increased from zero and exceeds a critical value $\alpha_{c}$, the participation number approaches asymptotically to the theoretical value for the delocalized eigenstates in a periodic lattice, $P=2N/3$ \cite{Casati}. This value is illustrated by the straight dashed line in the inset of Figure~\ref{fig:d}. This behavior was explained by the fact that the long-range correlation reduces the degree of disorder and causes delocalized states to occur. In other words, there exists a correlation-induced localization-delocalization transition at the critical point $\alpha_{c}$. As indicated in the inset of Figure~\ref{fig:d}, the critical value $\alpha_{c}$ depends on the disorder strength $W$. This is fully consistent with the result of previous works \cite{Kaya1}.

\begin{figure}
\centering
\includegraphics[width=10cm]{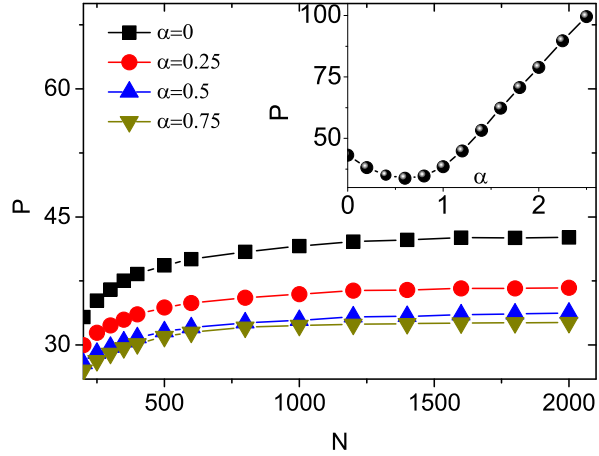}
\caption{Participation number $P$ versus system size $N$,
when the long-range correlated random on-site potential is imaginary-valued with zero mean.
The strength of disorder is fixed to $W=1.5$ and the correlation strength is chosen to be $\alpha=0$, 0.25, 0.5 and 0.75.
In this parameter region where $\alpha$ is small,
Anderson localization is enhanced due to the long-range correlation of the imaginary random potential.
In the inset, $P$ is plotted versus $\alpha$ when $W=1.5$ and $N=2000$.}
\label{fig:e}
\end{figure}

We now consider the case where the long-range correlated random on-site potential is imaginary-valued with zero mean.
In Figure~\ref{fig:e}, we plot $P$ versus $N$ for different values of the correlation strength $\alpha$ ($=0,0.25,0.5,0.75$),
when the disorder strength $W$ is fixed to 1.5. In this parameter region where $\alpha$ is small,
Anderson localization is enhanced due to the long-range correlation of the imaginary random potential.
Analogously to the short-range correlated case, this behavior is opposite to that shown in Figure~\ref{fig:d}, though the parameter values used in the two
calculations are similar.
As $\alpha$ increases further, however, $P$ attains a minimum and then increases again, as is shown in the inset. This non-monotonic dependence of
$P$ is similar to the short-range correlated case shown in Figure~\ref{fig:c}, though the increasing behavior beyond the minimum point is more steep
in the present case. Unlike in the case of a long-range correlated real random potential shown in Figure~\ref{fig:d}, we find that the eigenstates remain localized and no localization-delocalization transition occurs up to $\alpha=2.5$.

\begin{figure}
\centering
\includegraphics[width=10cm]{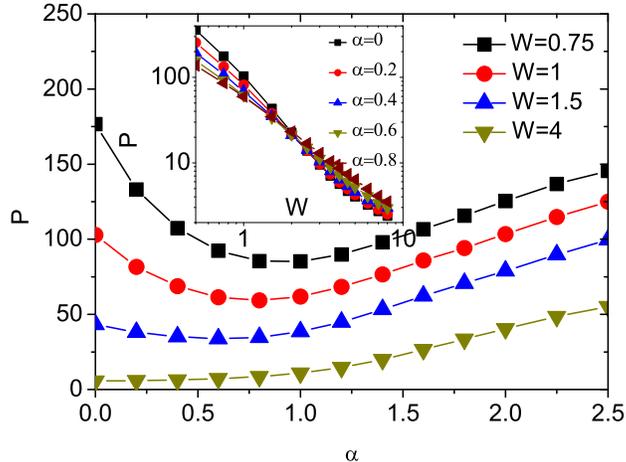}
\caption{Participation number $P$ versus correlation strength $\alpha$,
when the long-range correlated random on-site potential is imaginary-valued with zero mean.
The system size is fixed to $N=2000$ and the strength of disorder is chosen to be $W=0.75$, 1, 1.5 and 4.
In the weak correlation regime where $\alpha<1$, there exists a critical value $W_{c}\simeq 2$ above (below) which the localization is suppressed (enhanced) with increasing $\alpha$, as demonstrated in the inset.}
\label{fig:f}
\end{figure}

In Figure~\ref{fig:f}, we show $P$ versus the correlation strength $\alpha$ for several different values of the disorder strength $W$
($=0.75,1,1.5,4)$, when the system size is fixed to $N=2000$. The behavior is rather similar to the short-range correlated case shown in Figure~\ref{fig:c}.
When $W$ is 0.75, 1 and 1.5, $P$ shows a non-monotonic dependence on $\alpha$, while when $W=4$, it increases monotonically
as $\alpha$ increases from 0. This implies that in the region where the correlation strength is sufficiently large,
the localization is suppressed due to long-range disorder correlations of the imaginary random potential. In the weak correlation regime where $\alpha<1$, there exists a critical value $W_{c}\simeq 2$ above (below) which the localization is suppressed (enhanced) with increasing $\alpha$, as demonstrated in the inset. We also find that in contrast to the case of a long-range correlated real random potential, there is no occurrence of localization-delocalization transition in the considered parameter regime.

\begin{figure}
\centering
\includegraphics[width=10cm]{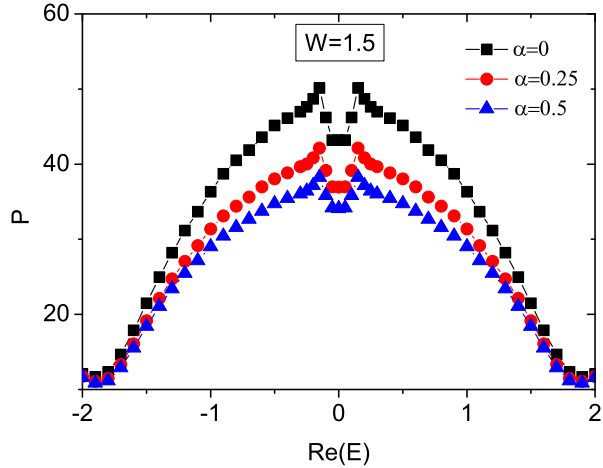}
\caption{Participation number $P$ versus real part of the energy eigenvalue ${\rm Re}(E)$,
when the long-range correlated random on-site potential is imaginary-valued with zero mean.
The system size and the disorder strength are fixed to $N=2000$ and $W=1.5$ ($<W_c$) and the correlation strength
is chosen to be $\alpha=0$, 0.25 and 0.5.
It is seen that the enhancement of localization occurs due to the long-range correlation of the imaginary random potential
in the whole energy band.}
\label{fig:g}
\end{figure}

In order to make sure that the obtained results do not arise from the peculiarity of the states in the vicinity of the band center, in Figure~\ref{fig:g},
we show the participation number $P$ as a function of the real part of the energy eigenvalue ${\rm Re}(E)$ for different values of the correlation exponent $\alpha$ ($=0,0.25,0.5$), when the disorder strength is fixed to $W=1.5$ ($<W_c$). We find that the enhancement of localization occurs due to the long-range correlation of the imaginary random potential in the whole energy band.
This enhancement of localization is much more pronounced in the vicinity of the band center than near the band edges. We observe that there appears
an anomalous non-analytic behavior of $P(E)$ in the vicinity of the band center for all $\alpha$.
This effect of anomalous localization enhancement near the band center has already been considered in details by us in a previous work \cite{Nguyen1}. We also observe a non-monotonic behavior of $P(E)$ near the band edges, which
can be understood in terms of Lifshitz tail states \cite{Kramer}. This kind of localized states originate from rare fluctuations of the on-site potential, where the values of the potential inside some sufficiently large volume turn out to be close to each other \cite{Silvestrov}.

\section{Conclusion}
\label{sec:con}

In this paper, we have presented a numerical investigation of the effects of short- and long-range disorder correlations on the localization properties of the eigenstates in a 1D disordered lattice characterized by a random non-Hermitian Hamiltonian,
where the imaginary part of the on-site potential is a correlated random function of the position. In particular, we have numerically calculated the participation number, which measures the degree of spatial extension or localization of eigenstates, as a function of other parameters.
In the case of short-range correlations and when the correlation length is sufficiently small,
we have found that there is a critical value of the disorder strength, below which
localization is enhanced and above which it is suppressed, as the correlation length increases.
In the region where the correlation length is larger, localization has been
found to be suppressed in all cases as the correlation length increases.
A qualitatively similar behavior has been obtained for long-range correlations as the disorder strength and the correlation exponent are changed.
We have observed no signature of localization-delocalization transition
in the presence of a long-range correlated imaginary random potential, unlike in the presence of a long-range correlated real random potential.
In the parameter region where localization is enhanced in the presence of long-range correlations,
we have observed that the enhancement occurs in the entire
energy band, but is strongest near the band center. We have also found that the anomalous localization enhancement effect occurs at the band center.
Devices such as random lasers utilize highly disordered gain materials.
Our results can provide a useful guideline for achieving correlation-enhanced localized states in such structures,
thereby enhancing the device efficiency.

\section*{Disclosure statement}

No potential conflict of interest was reported by the authors.

\section*{Funding}
This research is funded by Vietnam National Foundation for Science and Technology Development (NAFOSTED) under Grant No. 103.01-2018.05. It is also supported by a National Research Foundation of Korea Grant (NRF-2019R1F1A1059024) funded by the Korean Government.

\end{document}